\documentclass[letterpaper, 10 pt, conference, twoside]{ieeeconf}

\IEEEoverridecommandlockouts

\usepackage{amssymb}
\usepackage{amsmath}
\usepackage{color}
\usepackage{graphics}
\usepackage{graphicx}
\usepackage{multirow}
\usepackage{cite}
\usepackage{algorithm,algorithmic}
\usepackage{savesym}
\savesymbol{AND}
\usepackage{epstopdf}
\usepackage{subcaption}
\usepackage{float}
\usepackage{tikz}
\usetikzlibrary{angles,quotes}
\usetikzlibrary{calc}
\usetikzlibrary{arrows,decorations.markings}

\newtheorem{theorem}{Theorem}
\newtheorem{remark}{Remark}
\newtheorem{lemma}{Lemma}
\newtheorem{corollary}{Corollary}
\newtheorem{definition}{Definition}
\newtheorem{proposition}{Proposition}

\title{\LARGE \bf A data-driven method for computing polyhedral invariant sets of black-box switched linear systems}

\author{Zheming Wang and Rapha\"el M. Jungers
\thanks{The authors are with the ICTEAM Institute, UCLouvain, Louvain-la-Neuve,1348, Belgium. Email addresses:  zheming.wang@uclouvain.be (Zheming Wang), raphael.jungers@uclouvain.be (Rapha\"el M. Jungers)}
\thanks{Rapha\"el M. Jungers is a  FNRS honorary Research Associate. This project has received funding from the European Research Council (ERC) under the European Union's Horizon 2020 research and innovation programme under grant agreement No 864017 - L2C. Rapha\"el M. Jungers is also supported by the Walloon Region and the Innoviris Foundation.}}

\begin{document}
\maketitle

\begin{abstract}
In this paper, we consider the problem of invariant set computation for black-box switched linear systems using merely a finite set of observations of system trajectories. In particular, this paper focuses on polyhedral invariant sets. We propose a data-driven method based on the one step forward reachable set. For formal verification of the proposed method, we introduce the concepts of $\lambda$-contractive sets and almost-invariant sets for switched linear systems. The convexity-preserving property of switched linear systems allows us to conduct contraction analysis on the computed set and derive a probabilistic contraction property. In the spirit of non-convex scenario optimization, we also establish a chance-constrained guarantee on set invariance. The performance of our method is then illustrated by numerical examples.
\end{abstract}

\section{Introduction}
Switched linear systems consist of a finite set of linear dynamics (called modes) and a switching rule that indicates the current active mode of the system. They constitute an important family of hybrid systems. While the system is governed by linear dynamics dwelling in the same mode, the jump from one mode to another causes interesting hybrid phenomena distinct from the behaviors of the individual linear dynamics. For instance, despite the simplicity of the dynamics, stability analysis for a switched linear system is still complicated due to the switching signal, see \cite{ART:LA09} and the references therein.

Invariant set theory is widely used in system analysis and has been successfully generalized to study the properties of switched systems, see, e.g., \cite{BOO:BM08}. One typical technique for invariant set characterization is to construct Lyapunov functions of the switched system, see, e.g., \cite{ART:BMS07, ART:LA09, ART:RSD10}. In the presence of state constraints, more complications arise because invariant sets have to be constraint admissible, see \cite{ART:DO12} for the case of polyhedral constraints.  While handling general nonlinear constraints is still an open problem, there exist algorithms for computing invariant sets for certain classes of nonlinear constraints, see, e.g., \cite{INP:AJ16,ART:WJO19}. In \cite{ART:AJ18}, combinatorial methods have been introduced for switched systems where the switching signals are restricted by a labeled directed graph or an automaton. 

The aforementioned algorithms are all based on the knowledge of a hybrid model of the switched system, which is usually obtained by hybrid system
identification \cite{BOO:LB18}. However, except for simple systems with very low dimensions, hybrid system identification is often computationally demanding. In fact, identifying a switched linear system is known to be NP-hard \cite{ART:L16}. Data-driven analysis under the framework of black-box systems has been an active area of research in recent years, see \cite{INP:KQHT16,ART:KBJT19,ART:WJ20}. For instance, probabilistic stability guarantees are provided in \cite{ART:KBJT19} for black-box switched linear systems, based merely on a finite number of observations of trajectories. Data-driven analysis also allows us to study set invariance for black-box systems.  Recently, a scenario-based set invariance verification method has been proposed in \cite{ART:WJ20} for black-box discrete-time nonlinear systems. However, this method does not apply to switched systems with arbitrarily switching.  In this paper, we consider the computation of invariant sets of black-box switched linear systems under arbitrary switching. The data-driven stability analysis technique in \cite{ART:KBJT19} essentially attempts to compute an invariant ellipsoid. However, ellipsoidal invariant sets are often conservative for switched linear systems, because they rely on a common quadratic Lyapunov function, which may not exist even if the system is stable, see \cite{ART:LA09}.  Hence, we focus on polyhedral invariant sets of switched linear systems. More specifically, our goal is to develop a data-driven method for computing polyhedral invariant sets in the spirit of the scenario optimization approach \cite{ART:CGR18}. The contributions of this paper are threefold. First, inspired by \cite{ART:GP13}, we propose a geometric algorithm based on a finite set of snapshot pairs of the states.  Second, we introduce the concept of almost-invariant sets for switched linear systems and show their connections to \emph{$\lambda$-contractive} sets via contraction analysis. Third, we derive probabilistic guarantees for the set computed from the geometric algorithm.

The rest of the paper is organized as follows. This section ends with the notation, followed by the next section on the review of preliminary results on invariant sets and switched linear systems. Section \ref{sec:method} presents the proposed data-driven method. In Section \ref{sec:bound}, probabilistic guarantees of the proposed method are discussed. Numerical results are provided in Section \ref{sec:num}.

\textbf{Notation}. The non-negative integer set is indicated by $\mathbb{Z}^+$. For a square matrix $Q$, $Q \succ (\succeq) ~ 0$ means $Q$ is positive definite (semi-definite). $\mathbb{S}_{n-1}$ and $\mathbb{B}_n$ are the unit sphere and unit ball respectively in $\mathbb{R}^n$.  Let  $\mu(\cdot)$ denote the uniform spherical measure on $\mathbb{S}_{n-1}$ with $\mu(\mathbb{S}_{n-1}) = 1$. For any symmetric matrix $P\succ 0$, we define  $\|x\|_P:=\sqrt{x^TPx}$. Given any set $S\subseteq \mathbb{R}^n$, $\textrm{conv}(S)$ is the convex hull of $S$ and let $\|x\|_S$ denote $\min\{\lambda\ge 0: x\in \lambda S\}$ for any $x\in \mathbb{R}^n$.
A (bounded) polytope $S$ is called a C-polytope if it is convex and contains the origin in its interior. For any C-polytope $S$, let $\mathcal{V}(S)$ denote the set of  vertices and $\mathcal{F}(S)$ denote the set of facets. Given any $u\in \mathbb{R}^n$ and $\theta\in [0,\pi/2]$, let $Cap(u,\theta):=\{v\in \mathbb{S}_{n-1}: u^Tv \ge \|u\|\cos(\theta)\}$ denote the spherical cap with the direction $u$ and the angle $\theta$. 

\section{Preliminaries and problem statement}\label{sec:pre}
Switched linear systems are described below:
\begin{align}\label{eqn:Asigma}
x(t+1) = A_{\sigma(t)}x(t), \quad t\in \mathbb{Z}^+
\end{align}
where $\sigma(t): \mathbb{Z}^+\rightarrow \mathcal{M}:=\{1,2,\cdots,M\}$ a time-dependent switching signal that indicates the current active mode of the system among $M$ possible modes in $\mathcal{A}:=\{A_1,A_2,\cdots, A_M\}$. For any given switching sequence $\sigma$, let 
\begin{align}\label{eqn:pmbAsigmak}
\pmb{A}_{\pmb{\sigma}(k)} := A_{\sigma(k-1)} \cdots A_{\sigma(1)} A_{\sigma(0)}   , \quad k\in \mathbb{Z}^+
\end{align}
with $\pmb{\sigma}(k):=\{\sigma(k-1),\cdots, \sigma(1), \sigma(0)\}$, $\pmb{\sigma}(0) = \emptyset$, and $\pmb{A}_{\pmb{\sigma}(0)} = I_n$. The stability of System (\ref{eqn:Asigma}) can be described by the joint spectral radius (JSR) of the matrix set $\mathcal{A}$ defined by \cite{BOO:J09}
\begin{align}
\rho(\mathcal{A}): = \lim\limits_{k\rightarrow \infty} \max\limits_{\pmb{\sigma}(k)\in \mathcal{M}^k}\|\pmb{A}_{\pmb{\sigma}(k)}\|^{1/k}
\end{align}
Throughout the paper, we assume that $\rho(\mathcal{A})< 1$. We focus on the computation of invariant sets of System (\ref{eqn:Asigma}) under arbitrary switching, which are formally defined below.

\begin{definition}
A nonempty set $Z\subseteq \mathbb{R}^n$ is an invariant set for System (\ref{eqn:Asigma}) if $x\in Z$ implies that $Ax\in Z$ for any $A\in \mathcal{A}$. 
\end{definition}

From the definition above, invariant sets are inherently related with the stability of System (\ref{eqn:Asigma}). For instance, the level set of a common quadratic Lyapunov function,  which can be efficiently computed via semidefinite programming when it exists and the dynamics matrices $\mathcal{A}$ are known, see, e.g., \cite{ART:LA09}, is an ellipsoidal invariant set. In this paper, we focus on polyhedral invariant sets. Under the assumption that $\rho(\mathcal{A})<1$, the existence of a polyhedral invariant set is guaranteed,  while an ellipsoidal invariant set may not exist because a common quadratic Lyapunov function does not necessarily exist. This is one of the reasons why polyhedral invariant sets are often more appealing for switched linear systems, even though the computation may be more expensive.

A necessary and sufficient condition for set invariance in the polyhedral case is given below.
\begin{proposition}
A  C-polytope $S\subseteq \mathbb{R}^n$ is an invariant set for System (\ref{eqn:Asigma}) if and only if
\begin{align}\label{eqn:AxSxS}
\|A_{\sigma}x\|_{S} \le \|x\|_{S}, \quad \forall x\in \mathbb{S}_{n-1}, \forall \sigma\in \mathcal{M}.
\end{align}
\end{proposition}
Proof: This proposition is a direct consequence of the homogeneity property, i.e., for any $\gamma >0$, $\|\gamma x\|_{S} = \gamma \|x\|_{S}$ and $\|A_{\sigma}\gamma x\|_{S} = \gamma \|A_{\sigma}x\|_{S}$.  $\Box$

When the dynamics matrices $\mathcal{A}$ are known, classical algorithms based on iterative linear programming exist, see, e.g., \cite{ART:DO12,ART:GP13},  allowing to compute such a set efficiently. However, as we have mentioned above, in many cases, approximating the model of a switched system is computationally demanding, let alone identifying it exactly.  This paper considers the case where the dynamics matrices $\mathcal{A}$ are unknown. We call such systems black-box switched linear systems. 

In the black-box case, we sample a finite set of the initial states and the switching modes. More precisely, we randomly and uniformly generate $N$ initial states on $\mathbb{S}_{n-1}$ and $N$ modes in $\mathcal{M}$, which are denoted by $\omega_N:=\{(x_i,\sigma_i)\in \mathbb{S}_{n-1} \times \mathcal{M}: i=1,2,\cdots,N\}$.  From this random sampling, we observe the data set $\{(x_i,A_{\sigma_i}x_i): i=1,2,\cdots,N\}$, where $A_{\sigma_i}x_i$ is the successor of the initial state $x_i$. Note that the switching signal does not have to be observable.

For the given data set $\omega_N$ (or $\{(x_i,A_{\sigma_i}x_i)\}_{i=1}^N$), we define the following sampled problem:
\begin{align}\label{eqn:AxSxSomega}
\textrm{find } S 
\textrm{ s.t. }\|A_\sigma x\|_{S} \le \|x\|_{S}, \forall (x,\sigma) \in \omega_N
\end{align}
where $S$ is a C-polytope. As we assume asymptotic stability under arbitrary switching, we are interested in invariant sets that contain the origin in their interiors. For this reason, $S$ in (\ref{eqn:AxSxSomega}) is  restricted to be a C-polytope. In this paper, we attempt to solve this sampled problem (\ref{eqn:AxSxSomega}) using a geometric algorithm by scaling the sampled points and computing the convex hull of the scaled points iteratively. We will show that convergence of this algorithm is guaranteed under the assumption that $\rho(\mathcal{A})<1$.

\section{Data-driven computation of polyhedral invariant sets }\label{sec:method}
This section presents the proposed data-driven method for computing polyhedral invariant sets of black-box switched linear systems.

\subsection{A geometric algorithm}
We first present a geometric algorithm for computing invariant sets for the case where the matrices $\mathcal{A}$ are known. This geometric algorithm is based on the one step forward reachable set \cite{BOO:BM08,ART:GP13}. Given an initial C-polytope $X$, let us define:
\begin{align}
R_{k+1} = \textrm{conv}(R_k \bigcup\limits_{\sigma\in \mathcal{M}} A_{\sigma}R_k), R_0 &= X,  k\in \mathbb{Z}^+. \label{eqn:Rk}
\end{align}
The properties of the algorithm above are stated in the following proposition. 
\begin{proposition}[\cite{ART:GP13}]\label{prop:mininv}
Suppose $\rho(\mathcal{A})<1$, let us define $R_{k}$ as in (\ref{eqn:Rk}) for all $k\in \mathbb{Z}^+$ with an initial C-polytope $X$. Then, the following results hold. 
(i) There exists a finite $k$ such that $R_{k+1} =R_{k} = R_{\infty}$.
(ii) The set $R_{\infty}$ is the smallest invariant set that contains $X$.
\end{proposition}
Proof: A sketch of the proof is given here. We refer the readers to \cite{ART:GP13} for the detailed proof. From (\ref{algo:datainv}), $\forall k\in \mathbb{Z}^+$,
$
R_{k} =  \textrm{conv}(X\bigcup_{\sigma\in \mathcal{M}} A_{\sigma}X \bigcup \cdots \bigcup_{\pmb{\sigma}\in \mathcal{M}^k} \pmb{A}_{\pmb{\sigma}} X)
$
where $\pmb{A}_{\pmb{\sigma}}$ is defined in (\ref{eqn:pmbAsigmak}). Since $\rho(\mathcal{A})<1$ and $X$ is a C-polytope, there always exists a $k$ such that  $\pmb{A}_{\pmb{\sigma}} X \subseteq X$ for all $\pmb{\sigma} \in \mathcal{M}^{k+1}$, which implies that $R_{k+1} = R_k=R_{\infty}$. 
$\Box$

\subsection{The proposed data-driven method}\label{sec:iter}
With the sample $\omega_N$ and an initial C-polytope $X$, we now present a data-driven version of the geometric algorithm (\ref{eqn:Rk}):
\begin{align}
\tilde{R}_{k+1}(\omega_N) = \textrm{conv}(\tilde{R}_{k}(\omega_N) \cup \Omega_k(\omega_N) ), \quad \forall k\in \mathbb{Z}^+ \label{eqn:tildeRk}
\end{align}
where $\tilde{R}_0(\omega_N) = X$ and 
\begin{align}\label{eqn:Omegak}
\Omega_k(\omega_N) :=& \{\frac{A_\sigma x}{\|x\|_{\tilde{R}_k(\omega_N)}}: (x,\sigma) \in \omega_N \} \nonumber \\
&\cup \{\frac{-A_\sigma x}{\|-x\|_{\tilde{R}_k(\omega_N)}}: (x,\sigma) \in \omega_N \} .
\end{align}
The convergence of the data-driven geometric algorithm is stated in the following lemma.
\begin{theorem}\label{thm:mindata}
Suppose $\rho(\mathcal{A})<1$. Given a sample of $N$ points in $\mathbb{S}_{n-1}\times \mathcal{M}$, denoted by $\omega_N$, let $R_{k}$ and $\tilde{R}_k(\omega_N)$ be defined as in (\ref{eqn:Rk})  and (\ref{eqn:tildeRk}) respectively for all $k\in \mathbb{Z}^+$ with the same initial C-polytope $X$. Then, the following results hold.
(i) For any $k\in \mathbb{Z}^+$, $\tilde{R}_{k}(\omega_N)\subseteq R_k$.
(ii) The sequence $\{\tilde{R}_{k}(\omega_N)\}_{k\in \mathbb{Z}^+}$ is convergent.
(iii) $\tilde{R}_{\infty}(\omega_N)$ is a feasible solution to Problem (\ref{eqn:AxSxSomega}).
\end{theorem}
Proof: (i) The proof goes by induction. Suppose $\tilde{R}_{k}(\omega_N)\subseteq R_k$ for some $k\in \mathbb{Z}^+$. From the definition in (\ref{eqn:Omegak}), it holds that $\Omega_k(\omega_N) \subseteq \bigcup_{\sigma\in \mathcal{M}} A_{\sigma}R_k$. Hence, $\tilde{R}_{k+1}(\omega_N)\subseteq R_{k+1}$. Thus, the statement is true as $\tilde{R}_{0}(\omega_N)\subseteq R_0$.
(ii) The convergence of $\{\tilde{R}_{k}(\omega_N)\}_{k\in \mathbb{Z}^+}$ is a direct consequence of (i). 
(iii) From (\ref{eqn:tildeRk}), when $\{\tilde{R}_{k}(\omega_N)\}_{k\in \mathbb{Z}^+}$ converges, $\Omega_{\infty}(\omega_N)  \subseteq \tilde{R}_{\infty}(\omega_N)$, which implies that $\frac{A_\sigma x}{\|x\|_{\tilde{R}_{\infty}(\omega_N)}}\in \tilde{R}_{\infty}(\omega_N)$ for any $(x,\sigma)\in \omega_N$. Hence, $\|A_{\sigma}x\|_{\tilde{R}_{\infty}(\omega_N)} \le \|x\|_{\tilde{R}_{\infty}(\omega_N)}$ for any $(x,\sigma)\in \omega_N$. $\Box$

The theorem above shows that  $\{\tilde{R}_{k}(\omega_N)\}_{k\in \mathbb{Z}^+}$ eventually converges to a feasible solution of the sampled problem  (\ref{eqn:AxSxSomega}). However, finite-time convergence of (\ref{eqn:Rk}) may not be preserved. For the practical implementation, we use a stopping criterion as shown in Algorithm \ref{algo:datainv}.

\begin{algorithm}[h]
\caption{Data-driven computation of polyhedral invariant sets}
\begin{algorithmic}[1]
\renewcommand{\algorithmicrequire}{\textbf{Input:}}
\renewcommand{\algorithmicensure}{\textbf{Output:}}
\REQUIRE $X$, $\omega_N$ and some tolerance $\epsilon >0$
\ENSURE $\tilde{R}_k(\omega_N)$\\
\textit{Initialization}: Let $k\leftarrow 0$ and $\tilde{R}_k(\omega_N) \leftarrow X$;
\STATE Obtain $\Omega_k(\omega_N)$ from (\ref{eqn:Omegak});
\IF {$\Omega_k(\omega_N) \subseteq (1+\epsilon) \tilde{R}_k(\omega_N)$}
\STATE Terminate;
\ELSE
\STATE Compute  $\tilde{R}_{k+1}(\omega_N)$ from (\ref{eqn:tildeRk});
\STATE Let $k\leftarrow k+1$ and go to Step 1.
\ENDIF
\end{algorithmic}
\label{algo:datainv}
\end{algorithm}


\section{Probabilistic set invariance guarantees}\label{sec:bound}
In this section, we formally discuss probabilistic guarantees on the data-driven method proposed in Section \ref{sec:method}.

\subsection{$\lambda$-contractive sets and almost-invariant sets}
While the polyhedral set obtained from Algorithm \ref{algo:datainv} may not be an exact invariant set because we only use a finite number of samples, it does enjoy some properties in the probabilistic sense. To state these properties, we need to bring in two concepts for switched linear systems. Let us first recall the concept of \emph{$\lambda$-contractive} sets.
\begin{definition}
Given $\lambda \ge 0$, a set $S\subseteq \mathbb{R}^n$ is a \emph{$\lambda$-contractive} set for System (\ref{eqn:Asigma}) if $x\in S$ implies that $Ax\in \lambda S$ for any $A\in \mathcal{A}$. When $\lambda > 1$, the set can be in fact expansive, but we still call it a \emph{$\lambda$-contractive} set to be consistent.
\end{definition}

We also generalize the definition of invariant sets
for black-box switched linear systems. Here, we consider almost-invariant sets, which refer to sets that are invariant almost everywhere except in an small subset, see the definition below.

\begin{definition}[adapted from \cite{ART:DJ99,ART:WJ20}]\label{def:inv}
Given $\epsilon\in (0,1]$, a set $S\subseteq \mathbb{R}^n$ is an $\epsilon$ almost-invariant set for System (\ref{eqn:Asigma}) if $
\mu(\{x\in \mathbb{S}_{n-1}:\|Ax\|_{S} \le \|x\|_{S}, ~ \forall A\in \mathcal{A}\}) \ge 1-\epsilon$, where $\mu(\cdot)$ denotes the uniform spherical measure.
\end{definition}

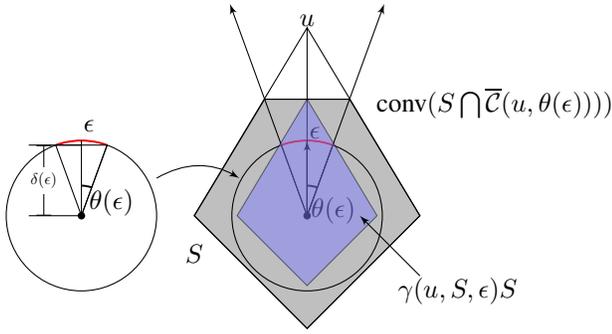
\begin{figure}[h]
\tikzset{>=latex'}
\def\myrad{1cm}
\def\myang{40}
\centering
\begin{tikzpicture}
\coordinate (O) at (0,0);

\draw[fill=gray!50] (-1.5,0) -- (0,-1.5) -- (1.5,0) -- (70:1.65) -- (110:1.65) -- cycle;

\draw (O) node[circle,inner sep=1.0pt,fill] {} circle [radius=\myrad];
\draw 
  (70:\myrad ) coordinate (xcoord) -- 
  node[midway,below] {} (O) -- 
  (110:\myrad) coordinate (slcoord)
  pic [draw,color=red, line width=0.25mm,angle radius=\myrad] {angle = xcoord--O--slcoord};
\draw 
  (70:1 ) coordinate (xcoord) -- 
  node[midway,below] {} (O) -- 
  (90:1) coordinate (slcoord)
  pic [draw,color=black, line width=0.25mm,angle radius=0.4*\myrad] {angle = xcoord--O--slcoord};

\draw (10pt,2pt) node {$\theta(\epsilon)$};
\draw (3pt,1.1) node {$\epsilon$};

\draw (O) -- (0,2.5) -- (-1.5,0) -- (0,-1.5) -- (1.5,0) -- (0,2.5);
\draw [->] (O) -- (70:3);
\draw [->] (O) -- (110:3);
\draw (70:1.65) -- (110:1.65);

\draw (-1.5,-0.5) node{$S$};
\draw (2.5,1.5) node{$\textrm{conv}(S \bigcap \overline{\mathcal{C}}(u,\theta(\epsilon))))$};

\draw [->] (O) -- (0,1);
\draw (0,2.6) node {$u$};
\draw[fill=blue!50,opacity=0.5] (0,2.5*0.6202) -- (-1.5*0.6202,0) -- (0,-1.5*0.6202) -- (1.5*0.6202,0) -- (0,2.5*0.6202);
\draw [->] (1.5,-0.8) -- ++ (-0.8,0.8);
\draw (2,-1) node{$\gamma(u,S,\epsilon)S$};
\draw (-2,0.5) edge[bend left,->] (-0.9,0.5) ;

\coordinate (Ol) at (-3,0);
\draw (Ol) node[circle,inner sep=1.0pt,fill] {} circle [radius=\myrad];
\draw 
  (70:\myrad)++(Ol) coordinate (xcoord) --  (Ol) -- ++
  (110:\myrad) coordinate (slcoord)
  pic [draw,color=red, line width=0.25mm,angle radius=\myrad] {angle = xcoord--Ol--slcoord};

\draw 
  (70:1 )++(Ol) coordinate (xcoord) --  (Ol) -- ++
  (90:1) coordinate (slcoord)
  pic [draw,color=black, line width=0.25mm,angle radius=0.4*\myrad] {angle = xcoord--Ol--slcoord};

\draw (-3+0.3420,0.9397) -- (-3-0.3420,0.9397);
\draw (-3+0.4,5pt) node {$\theta(\epsilon)$};
\draw (-3+0.1,1.2) node {$\epsilon$};
\draw (Ol) -- ++(-0.7,0) ;
\draw (-3,0.9397) -- ++(-0.7,0) ;
\draw [|-|](-3.5,0.9397) -- (-3.5,0) node[midway,fill=white,scale=0.6] {$\delta(\epsilon)$} ;
\end{tikzpicture}
\caption{Geometric illustration of the definition $\gamma(u,S,\epsilon)$: the red curve denotes the subset of measure $\epsilon$ on  the unit sphere, the gray area denotes $\textrm{conv}(S \bigcap \overline{\mathcal{C}}(u,\theta(\epsilon))))$ and the blue area denotes $\gamma(u,S,\epsilon)S$. }\label{fig:ill}
\end{figure}

We then show that an $\epsilon$ almost-invariant set for System (\ref{eqn:Asigma}) is also a \emph{$\lambda$-contractive} set for some $\lambda >0$. To obtain a tight contraction rate, we introduce additional definitions as follows. For any $\epsilon\in (0,1/2)$, let
\begin{align}
\delta(\epsilon) &:= \sqrt{1-\mathcal{I}^{-1}(2\epsilon;\frac{n-1}{2},\frac{1}{2})}, \label{eqn:deltaep}\\
\theta(\epsilon) &:= \cos^{-1}(\delta(\epsilon)), \label{eqn:thetaep} 
\end{align}
where $\mathcal{I}(x;a,b)$ is the regularized incomplete beta function (see, e.g., \cite{ART:KBJT19}) defined as
\begin{align}
\mathcal{I}(x;a,b) := \frac{\int_0^x t^{a-1}(1-t)^{b-1}dt}{\int_0^1 t^{a-1}(1-t)^{b-1}dt}.
\end{align}
For any given C-polytope $S\subseteq \mathbb{R}^n$ and $u\in \mathcal{V}(S)$, let
\begin{align}\label{eqn:gammauSep}
\gamma(u,S,\epsilon) &:= \max_{\alpha\ge 0} \{\alpha:  \alpha u \in \textrm{conv}(S \bigcap \overline{\mathcal{C}}(u,\theta(\epsilon))) \}
\end{align}
where $\epsilon\in (0,1/2)$ and $\overline{\mathcal{C}}(u,\theta)$ is given by
\begin{align}
\overline{\mathcal{C}}(u,\theta) := \{x\in \mathbb{R}^n: u^Tx \le \|x\| \|u\|\cos(\theta)\}, 
\end{align}
which is the closure of the complement of the cone $\mathcal{C}(u,\theta)$ with the direction $u$ and the angle $\theta$:
\begin{align}\label{eqn:Cutheta}
\mathcal{C}(u,\theta) := \{x\in \mathbb{R}^n: u^Tx \ge \|x\|\|u\|\cos(\theta)\}.
\end{align}
A geometric illustration of the definition in (\ref{eqn:gammauSep}) is illustrated in Figure \ref{fig:ill}. Let us define:
\begin{align}\label{eqn:gammamin}
\gamma_{\min}(S,\epsilon) := \min_{u\in \mathcal{V}(S)} \gamma(u,S,\epsilon).
\end{align} 

With these definitions, we state the contraction property of almost-invariant sets in the following proposition.

\begin{proposition}\label{prop:contraction}
Given $\epsilon\in (0,1/2)$, suppose a C-polytope $S\subseteq \mathbb{R}^n$ is an $\epsilon$ almost-invariant set of System (\ref{eqn:Asigma}). Let $\gamma_{\min}(S,\epsilon)$ be defined as in (\ref{eqn:gammamin}). Then, $S$ is a \emph{$\lambda$-contractive} set of System (\ref{eqn:Asigma}) with $\lambda = 1/\gamma_{\min}(S,\epsilon)$.
\end{proposition}

The proof of Proposition \ref{prop:contraction} is given in the appendix. From the definition in (\ref{eqn:gammamin}), to obtain $\gamma_{\min}(S,\epsilon)$, we need to compute $\gamma(u,S,\epsilon)$, which requires the computation of  
$
 \textrm{conv}(\{x\in S: u^Tx \le \|x\| \|u\| \cos(\theta)\})
$
for all $u\in \mathcal{V}(S)$. In general, computing the convex hull of a nonlinear constraint set is a difficult problem, see \cite{ART:NRT18}. For this reason, we formulate a relaxation of (\ref{eqn:gammamin}), which yields a computational tractable lower bound which can be computed by solving convex problems. Given a C-polytope $S$ and $\epsilon \in (0,1/2)$, we define:
\begin{align}\label{eqn:gammaunder}
\underline{\gamma}(S,\epsilon) &: = \min_{u\in \mathcal{V}(S)}  \delta(\epsilon)d_{\min}(u,S,\epsilon)/\|u\|
\end{align}
where $\delta(\epsilon)$ is defined in (\ref{eqn:deltaep}), and 
\begin{align}\label{eqn:dminvS}
d_{\min}(u,S,\epsilon) : = \min_{x\in \partial S} \{\|x\| : x\in \mathcal{C}(u,\theta(\epsilon))\}.
\end{align}
The properties of $\underline{\gamma}(S,\epsilon)$ are given in the following lemma.

\begin{lemma}\label{lem:bound}
Given any C-polytope $S\subseteq \mathbb{R}^n$ and any $\epsilon\in (0,1/2)$, let us define $\gamma_{\min}(S,\epsilon)$ and $\underline{\gamma}(S,\epsilon)$ in (\ref{eqn:gammamin}) and (\ref{eqn:gammaunder}) respectively. Then, $\gamma_{\min}(S,\epsilon) \ge \underline{\gamma}(S,\epsilon)$.
\end{lemma}
The proof of Lemma \ref{lem:bound} is given in the appendix. We then show that $\underline{\gamma}(S,\epsilon)$ can be computed by solving a set of convex optimization problems. Given any C-polytope, let us define the following problem, for any $u\in \mathcal{V}(S)$ and $f\in \mathcal{F}(S)$ that satisfy $f\cap \mathcal{C}(u,\theta(\epsilon)) \not=\emptyset$, 
\begin{align}\label{eqn:dminfvS}
d_{\min}^f(u,S,\epsilon) &:=  \min_{x\in f} \{\|x\|: u^Tx \ge \delta(\epsilon)\|x\|\|u\|\}
\end{align}
This is a convex problem and can be efficiently solved by classic solvers, see \cite{BOO:BV04}. The following lemma shows that $\underline{\gamma}(S,\epsilon)$ defined in (\ref{eqn:dminvS}) can be computed by solving (\ref{eqn:dminfvS}).

\begin{lemma}\label{lem:com}
Given any $\epsilon\in (0,1/2)$, C-polytope $S\subseteq \mathbb{R}^n$, and $u\in \mathcal{V}(S)$,  one has: 
\begin{align}
d_{\min}(u,S,\epsilon) = \min_{f\in \mathcal{F}(S)} d_{\min}^f(u,S,\epsilon),
\end{align}
where $d_{\min}(u,S,\epsilon)$ and $d_{\min}^f(u,S,\epsilon)$ are defined in (\ref{eqn:dminvS}) and (\ref{eqn:dminfvS}) respectively.
\end{lemma}
Proof: To compute $d_{\min}(u,S,\epsilon)$, we need to check all the points on $\partial S \cap \mathcal{C}(u,\theta(\epsilon))$. This can be equivalently done by checking all the facets of $S$ and solving Problem (\ref{eqn:dminfvS}). $\Box$

\begin{remark}\label{rem:disjoint}
Suppose $\tilde{\mathbb{S}} = \{x\in \mathbb{S}_{n-1}:\|Ax\|_{S} \le \|x\|_{S}, ~ \forall A\in \mathcal{A}\}$. From the proof of Proposition \ref{prop:contraction}, the results above also hold for the case where the violating subset $\mathbb{S}_{n-1} \setminus \tilde{\mathbb{S}}$ is the union of a group of disjoint sets whose measures are bounded by $\epsilon$.
\end{remark}

\subsection{Contraction analysis}
With the discussion above, we are now in a position to derive a probabilistic contraction property of the set computed from Algorithm \ref{algo:datainv}. Let us recall the notions of covering and packing numbers, see, e.g., Chapter 27 of \cite{BOO:SB14}.
\begin{definition}
Given $\epsilon\in (0,1/2)$, a set $U \subset \mathbb{S}_{n-1}$ is called an \emph{$\epsilon$-covering} of $\mathbb{S}_{n-1}$ if, for any $x\in \mathbb{S}_{n-1}$, there exists $u\in U$ such that $u^T x \ge \delta(\epsilon)$. The \emph{covering number} $\mathcal{N}_c(\epsilon)$ is the minimal cardinality of an \emph{$\epsilon$-covering} of $\mathbb{S}_{n-1}$.
\end{definition}

\begin{definition}
Given $\epsilon\in (0,1/2)$, a set $U \subset \mathbb{S}_{n-1}$ is called an \emph{$\epsilon$-packing} of $\mathbb{S}_{n-1}$ if, for any $u,v\in U$,  $u^T v > \delta(\epsilon)$. The \emph{packing number} $\mathcal{N}_p(\epsilon)$ is the maximal cardinality of an \emph{$\epsilon$-packing} of $\mathbb{S}_{n-1}$.
\end{definition}

With these two notions, the following lemma is obtained.

\begin{lemma}\label{lem:covering}
For any $\epsilon\in (0,1/2)$, let $\delta(\epsilon)$ and  $\theta(\epsilon)$ be defined in (\ref{eqn:deltaep}) and (\ref{eqn:thetaep}) respectively. Then,
\begin{align}
\mathcal{N}_c(\epsilon) \le \mathcal{N}_p(\epsilon) \le \frac{2}{\mathcal{I}(\sin^2(\frac{\theta(\epsilon)}{2});\frac{n-1}{2},\frac{1}{2})}.
\end{align}
\end{lemma}
Proof: Suppose $U$ is the \emph{$\epsilon$-packing} with the maximal cardinality. The first inequality follows from the fact that $U$ is also a \emph{$\epsilon$-covering}. For any direction $u\in \mathbb{S}_{n-1}$ and any angle $\theta\in [0,\pi/2]$, the spherical cap $Cap(u,\theta)$ has a measure of $\frac{1}{2}\mathcal{I}(\sin^2(\theta);\frac{n-1}{2},\frac{1}{2})$ (see \cite{ART:KBJT19} for details). From the definition of an \emph{$\epsilon$-packing}, the spherical caps $\{Cap(u,\theta(\epsilon)/2)\}_{u\in U}$ are disjoint. Hence, 
$
\sum_{u\in U} \mu\left(Cap(u,\theta(\epsilon)/2) \right)  \le 1
$, which leads to the second inequality. $\Box$

The probabilistic guarantee on contraction is then stated in the following theorem. Recall that $M$ is the number of modes in $\mathcal{M}$ (or $\mathcal{A}$).

\begin{theorem}\label{thm:polyinvconim}
Suppose $\rho(\mathcal{A})<1$. Given $N\in \mathbb{Z}^+$, let $\omega_N$ be i.i.d. with respect to the uniform distribution $\mathbb{P}$ over $\mathbb{S}_{n-1}\times \mathcal{M}$. With an initial C-polytope $X$, the set $\tilde{R}_{\infty}(\omega_N)$ is defined as in (\ref{eqn:tildeRk}). For any $\epsilon \in (0,1/2)$, let 
\begin{align}
\mathcal{B}(\epsilon;N) = \frac{2M(1-\frac{\epsilon}{M})^N}{\mathcal{I}(\sin^2(\frac{\theta(\epsilon)}{2});\frac{n-1}{2},\frac{1}{2})}.
\end{align}
where $\theta(\epsilon)$ is defined in (\ref{eqn:thetaep}). Then, given any $\epsilon \in (0,1/2)$, with probability no smaller than $1-\mathcal{B}(\epsilon;N)$, $\tilde{R}_{\infty}(\omega_N)$ is a \emph{$\lambda$-contractive} set of System (\ref{eqn:Asigma}) with $\lambda = 1/\underline{\gamma}(\tilde{R}_{\infty}(\omega_N),\mathcal{I}(\sin^2(2\theta(\epsilon));\frac{n-1}{2},\frac{1}{2})/2)$, where $\underline{\gamma}(\cdot,\cdot)$ is defined in (\ref{eqn:gammaunder}).
\end{theorem}
Proof: Consider the maximal \emph{$\epsilon$-packing} $U$ with the cardinality $\mathcal{N}_p$. From Lemma \ref{lem:covering}, $\{Cap(u,\theta(\epsilon))\}_{u\in U}$ covers $\mathbb{S}_{n-1}$.  Suppose $\omega_N$ is sampled randomly according to the uniform distribution, then the probability that each spherical cap in $\{Cap(u,\theta(\epsilon))\}_{u\in U}$ contains $M$ points with $M$ different modes is no smaller than $1-\mathcal{N}_pM(1-\frac{\epsilon}{M})^N \ge \mathcal{B}(\epsilon;N)$.  Hence, the angle of the largest spherical cap that violates the condition $\|Ax\|_{\tilde{R}_{\infty}(\omega_N)} \le \|x\|_{\tilde{R}_{\infty}(\omega_N)}, ~ \forall A\in \mathcal{A}$, is bounded by $2\theta(\epsilon)$. Thus, the measure of the largest violating spherical cap is bounded by $\mathcal{I}(\sin^2(2\theta(\epsilon));\frac{n-1}{2},\frac{1}{2})/2$. From Proposition \ref{prop:contraction}, and Lemmas \ref{lem:bound} \& \ref{lem:com}, $\tilde{R}_{\infty}(\omega_N)$ is a \emph{$\lambda$-contractive} set with the rate of $1/\underline{\gamma}(\tilde{R}_{\infty}(\omega_N),\mathcal{I}(\sin^2(2\theta(\epsilon));\frac{n-1}{2},\frac{1}{2})/2)$. $\Box$

\begin{remark}
As the dimension increases, the number of vertices of $\tilde{R}_{\infty}(\omega_N)$ increases. The computation of $\underline{\gamma}(\tilde{R}_{\infty}(\omega_N),M\varepsilon(s(\omega_N)))$ constitutes the main computational burden of our method.
\end{remark}


\subsection{Chance-constrained set invariance guarantee}
In the rest of this section, we show that the so-called chance-constrained theorem \cite{ART:CGR18} is applicable to the problem of invariant set computation of switched linear systems with the definition of $\epsilon$ almost-invariant sets in Definition \ref{def:inv}. To formally state the probabilistic guarantee on set invariance, we recall the definition of \emph{supporting points} in \cite{ART:CGR18}.

\begin{definition}[\cite{ART:CGR18}]\label{def:supp}
Consider a sample of $N$ points in $\mathbb{S}_{n-1}\times \mathcal{M}$, denoted by  $\omega_N$, and the iteration (\ref{eqn:tildeRk}) with an initial C-polytope $X$, $(x,\sigma)\in \omega_N$ is called a \emph{supporting point}, if $\tilde{R}_{\infty}(\omega_N\setminus \{(x,\sigma)\}) \not = \tilde{R}_{\infty}(\omega_N)$.  Let $s(\omega_N)$ denote the number of \emph{supporting points} in $\omega_N$. 
\end{definition}
The chance-constrained set invariance guarantee is stated in the following theorem.
\begin{theorem}[adapted from Theorem 1 in  \cite{ART:CGR18}]\label{thm:polyinv}
Suppose the same conditions as in Theorem \ref{thm:polyinvconim} hold. 
Let $\mathbb{P}^N$ denote the probability measure in the $N$-Cartesian product of $\mathbb{S}_{n-1}\times \mathcal{M}$. Let $\tilde{R}_{\infty}(\omega_N)$ be obtained from (\ref{eqn:tildeRk}) with an initial C-polytope $X$. Then, for any $\beta\in (0,1)$
\begin{align}\label{eqn:PNbeta}
\mathbb{P}^N(\{\omega_N:\mathbb{P}(V(\tilde{R}_{\infty}(\omega_N)))>\varepsilon(s(\omega_N))\}) \le \beta
\end{align}
where $V(\tilde{R}_{\infty}(\omega_N)):=\{(x,\sigma): \|A_{\sigma}x\|_{\tilde{R}_{\infty}(\omega_N)} > \|x\|_{\tilde{R}_{\infty}(\omega_N)}  \}$, $s(\omega_N)$ is the number of \emph{supporting points} as defined in Definition \ref{def:supp}, and $\varepsilon: \{0,1,\cdots, N\}\rightarrow [0,1]$ is a function defined as:
\begin{align}\label{eqn:varepsilonk}
\varepsilon(k):=\begin{cases} 
      1 & \textrm{if } k=N; \\
      1-\sqrt[N-k]{\frac{\beta}{N {N \choose k}}} & 0\le k<N.
   \end{cases}
\end{align} 
\end{theorem}

Since it is a simple adaptation of Theorem 1 in \cite{ART:CGR18}, the proof is omitted. Indeed, this bound is established a posteriori because it is based on the measured data $\omega_N$. With Theorem \ref{thm:polyinv} in hand, we can derive a probabilistic guarantee on set invariance in the following corollary. 
\begin{corollary}\label{cor:invset}
Suppose the same conditions as in Theorem \ref{thm:polyinv} hold. Let $\tilde{R}_{\infty}(\omega_N)$ be the solution obtained from (\ref{eqn:tildeRk}) with an initial C-polytope $X$ and $s(\omega_N)$ be defined in Definition \ref{def:supp}. Then, with probability no smaller than $1-\beta$, $\tilde{R}_{\infty}(\omega_N)$ is $M\varepsilon(s(\omega_N))$ almost-invariant set for System (\ref{eqn:Asigma}), where $\varepsilon(s(\omega_N))$ is defined in (\ref{eqn:varepsilonk}).
\end{corollary}
Proof: From Theorem \ref{thm:polyinv}, with probability no smaller than $1-\beta$, $ \mathbb{P}(V(\tilde{R}_{\infty}(\omega_N)))\le \varepsilon(s(\omega_N))$. Since $\{x\in \mathbb{S}_{n-1}:\|Ax\|_{\tilde{R}_{\infty}(\omega_N)} > \|x\|_{\tilde{R}_{\infty}(\omega_N)}\}=\{x\in \mathbb{S}_{n-1}: \exists \sigma\in \mathcal{M}:(x,\sigma)\in V(\tilde{R}_{\infty}(\omega_N)) \}$. Hence, $\mu(x\in \mathbb{S}_{n-1}:\|Ax\|_{\tilde{R}_{\infty}(\omega_N)} > \|x\|_{\tilde{R}_{\infty}(\omega_N)}) \le M \mathbb{P}(V(\tilde{R}_{\infty}(\omega_N))) \le M \varepsilon(s(\omega_N))$. $\Box$

From Proposition \ref{prop:contraction}, an $\epsilon$ almost-invariant set is also a \emph{$\lambda$-contractive} set for some $\lambda>0$. In this regard, the guarantee in Theorem \ref{thm:polyinv} (or Corollary \ref{cor:invset}) provides more information than the one in Theorem \ref{thm:polyinvconim}. However, as validated by numerical simulation in the next section, with the same number of samples and the same confidence level, the contraction rate obtained from Theorem \ref{thm:polyinvconim} is often better in most of cases (unless $s(\omega_N)$ is very small). In other words, Theorem \ref{thm:polyinv} provides more information with lower confidence level while Theorem \ref{thm:polyinvconim} provides less information with higher confidence level for the same setting.

\section{Numerical simulation}\label{sec:num}
We consider switched linear systems of dimension from $2$ to $8$ and number of modes from $4$ to $8$, generated using the JSR toolbox \cite{INP:VHJ14}. The initial set is $X=\{x: \|x\|_{\infty}\le 1\}$ and $10000$ points are sampled randomly and uniformly on the unit sphere. We then use Algorithm \ref{algo:datainv} to compute $\tilde{R}_{\infty}(\omega_N)$ with a tolerance of $10^{-8}$. While the matrices $\mathcal{A}$ are unknown, we still show $R_{\infty}$ for reference. In order to evaluate the difference between $R_{\infty}$ and its inner bound $\tilde{R}_{\infty}(\omega_N)$, we compute $\lambda^* = \max\{\lambda\ge 0: \lambda R_{\infty} \subseteq \tilde{R}_{\infty}(\omega_N) \}$. The results are given in Table \ref{tab:algo}, where $\tilde{k}$ is the number of iterations needed for Algorithm \ref{algo:datainv} and $k^*$ is the number of iterations needed for the standard algorithm in (\ref{eqn:Rk}). As expected, for low-dimensional systems, $\tilde{R}_{\infty}(\omega_N)$ can be considered as a good approximation of $R_{\infty}$, while, for high-dimensional systems, the difference is more significant.

\renewcommand{\arraystretch}{1.5}
\begin{table}[h]
\centering
\begin{tabular}{|c|c|c|c|c|c|}
\hline
$(n,M)$ & $\tilde{k}$ & $\mathcal{V}(\tilde{R}_{\infty}(\omega_N))$ & $k^*$ & $\mathcal{V}(R_{\infty})$ & $\lambda^*$ \\\hline
$(2,4)$ & $4$ & $8$ & $2$ & $8$ & $0.9992$ \\\hline
$(3,4)$ & $6$ & $20$ & $2$ & $22$ & $0.9551$ \\\hline
$(4,4)$ & $6$ & $34$ & $2$ & $38$ & $0.8863$ \\\hline
$(4,6)$ & $9$ & $52$ & $2$ & $48$ & $0.8795$ \\\hline
$(6,6)$ & $10$ & $192$ & $3$ & $272$ & $0.7093$ \\\hline
$(8,6)$ & $11$ & $322$ & $3$ & $1012$ & $0.6158$ \\\hline
$(8,8)$ & $11$ & $354$ & $3$ & $1196$ & $0.5837$ \\\hline
\end{tabular}
\caption{Performance of the proposed algorithm for different values of $n$ and $M$.}
\label{tab:algo}
\end{table}

\begin{figure}[h]
\centering
\includegraphics[width=1.9in]{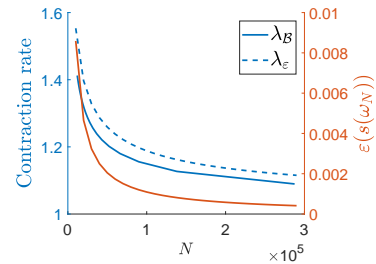}
\caption{Probabilistic guarantees obtained for the $3$-dimensional example with $4$ modes.}
\label{fig:bounds}
\end{figure}

For rigorous verification, we compute the probabilistic bounds derived in Theorems \ref{thm:polyinvconim} \& \ref{thm:polyinv} for the case where $n=3$ and $M=4$. We fix the confidence level at $\beta = 0.001$. For Theorem \ref{thm:polyinvconim}, we find $N$ such that $\mathcal{B}(\epsilon;N) = \beta$ for different values of $\epsilon$ and compute the contraction rate $\lambda_{\mathcal{B}}:=1/\underline{\gamma}(\tilde{R}_{\infty}(\omega_N),\mathcal{I}(\sin^2(2\theta(\epsilon));\frac{n-1}{2},\frac{1}{2})/2)$. Thus, we obtain a curve of contraction rate against the number of samples $N$. For Theorem \ref{thm:polyinv}, we compute $\varepsilon(s(\omega_N))$ and $\lambda_{\varepsilon}:=1/\underline{\gamma}(\tilde{R}_{\infty}(\omega_N),M\varepsilon(s(\omega_N)))$ for different values of $N$. The curves given in Figure \ref{fig:bounds} show that the contraction rate from Theorem \ref{thm:polyinvconim} is tighter.

\section{Conclusions}\label{sec:con}
We have presented a data-driven method for computing polyhedral invariant sets for black-box switched linear systems based on the one step forward reachable set. The convergence of this method is guaranteed under the stability assumption. Almost-invariant sets have been introduced for switched linear systems. The convexity-preserving property of switched linear systems allowed us to establish a probabilistic guarantee a priori via contraction analysis. With the chance-constraint theorem for nonconvex problems, we have also derived an a posteriori guarantee which provides a bound on the level of set invariance violation of the computed set. Finally, numerical examples are given to illustrate the performance of the proposed method.

\appendix
\textbf{Proof of Proposition \ref{prop:contraction}}: Let $\tilde{\mathbb{S}} = \{x\in \mathbb{S}_{n-1}:\|Ax\|_{S} \le \|x\|_{S}, ~ \forall A\in \mathcal{A}\}$ and 
$
\alpha^*:= \max_{\alpha\ge 0} \{\alpha : \alpha S \subseteq \textrm{conv}(\{x\in S:x/\|x\|\in \tilde{\mathbb{S}} \})\}
$.
For any $x\in \{x\in S:x/\|x\|\in \tilde{\mathbb{S}} \}$ and $A\in \mathcal{A}$, $Ax\in S$, which implies that $A\textrm{conv}(\{x\in S:x/\|x\|\in \tilde{\mathbb{S}} \}) \subseteq S$ for any $A\in \mathcal{A}$. Hence, $\alpha^*AS \subseteq A \textrm{conv}(\{x\in S:x/\|x\|\in \tilde{\mathbb{S}} \}) \subseteq S$ for any $A\in \mathcal{A}$. That is, for any $x\in S$, $Ax\in \frac{1}{\alpha^*} S$ for any $A\in \mathcal{A}$. Therefore, $S$ is a $\frac{1}{\alpha^*}$-contractive set. Now, it suffices to show that $\gamma_{\min}(S,\epsilon)$ is a lower bound of $\alpha^*$. For any $u\in \mathcal{V}(S)$, let $\bar{\alpha}(u):=\max_{\alpha\ge 0} \{\alpha : \alpha u \in  \textrm{conv}(\{x\in S:x/\|x\|\in \tilde{\mathbb{S}} \})\}$. Then, it holds that $\alpha^* = \min_{u\in \mathcal{V}(S)} \bar{\alpha}(u)$. In the rest of the proof, we aim to show that $\bar{\alpha}(u) \ge \gamma(u,S,\epsilon)$ for any $u\in \mathcal{V}(S)$. 
Suppose $\tilde{\theta}$ is the smallest $\theta$ such that the set $\{\alpha\ge 0 : \alpha u \in \textrm{conv}(\{x\in \partial S:x/\|x\|\in \tilde{\mathbb{S}} \cap Cap(u,\theta) \})\}$ is non-empty. Let $\tilde{\alpha}(u):=\max_{\alpha\ge 0} \{\alpha : \alpha u \in \textrm{conv}(\{x\in \partial S:x/\|x\|\in \tilde{\mathbb{S}} \cap Cap(u,\tilde{\theta}) \})\}$. Since $\textrm{conv}(\{x\in S:x/\|x\|\in \tilde{\mathbb{S}} \}) \supseteq \textrm{conv}(\{x\in \partial S:x/\|x\|\in \tilde{\mathbb{S}}  \cap  Cap(u,\tilde{\theta}) \})$, $\bar{\alpha}(u) \ge \tilde{\alpha}(u)$. The value of $\tilde{\alpha}(u)$ depends on the shape of  the violating set $\mathbb{S}_{n-1}\setminus \tilde{\mathbb{S}}$. Note that the set $(\mathbb{S}_{n-1}\setminus \tilde{\mathbb{S}})\setminus Cap(u,\tilde{\theta})$ does not affect the value of $\tilde{\alpha}(u)$. From this observation and the relation between the angel and the measure of the spherical cap (see \cite{ART:KBJT19} for details), we can see that $\tilde{\alpha}(u)$ reaches the minimal when $Cap(u,\tilde{\theta}) = \mathbb{S}_{n-1}\setminus \tilde{\mathbb{S}}$ with $\tilde{\theta}=\delta(\epsilon)$ as defined in (\ref{eqn:deltaep}). It can be verified that $\tilde{\alpha}(u)$ becomes $\gamma(u,S,\epsilon)$ in this case. Therefore, $\bar{\alpha}(u) \ge \gamma(u,S,\epsilon)$ and thus $\alpha^* \ge \gamma_{\min}(S,\epsilon)$. $\Box$

\textbf{Proof of Lemma \ref{lem:bound}}: (i) Since $\theta(\epsilon)\in (0,\pi/2)$, $\textrm{conv}(S \cap \overline{\mathcal{C}}(u,\theta(\epsilon))))=\textrm{conv}(\partial S \cap \overline{\mathcal{C}}(u,\theta(\epsilon))))$ for any $u\in \mathcal{V}(S)$. It is obvious that
$
\partial S = \left(\partial S \cap \overline{\mathcal{C}}(u,\theta(\epsilon))) \right) \cup \left(\partial S \cap \mathcal{C}(u,\theta(\epsilon)))\right).
$
Taking convex hull of both sides yields
$
S \subseteq \textrm{conv}(\partial S \cap \overline{\mathcal{C}}(u,\theta(\epsilon))) ) \cup \textrm{conv}(\partial S \cap \mathcal{C}(u,\theta(\epsilon)))),
$
which implies that
$
 \textrm{conv}(S \cap \overline{\mathcal{C}}(u,\theta(\epsilon)))) \supseteq S\setminus \textrm{conv}(\partial S \cap \mathcal{C}(u,\theta(\epsilon)))).
$
Thus,
\begin{align*}
\gamma(u,S,\epsilon) &\le \sup_{0 \le \alpha\le 1} \{\alpha:  \alpha u \in S\setminus \textrm{conv}(\partial S \bigcap \mathcal{C}(u,\theta(\epsilon)))) \}\\
&=\sup_{0 \le \alpha \le 1} \{\alpha:  \alpha u \not\in  \textrm{conv}(\partial S \bigcap \mathcal{C}(u,\theta(\epsilon)))) \}.
\end{align*}
From the definition in (\ref{eqn:Cutheta}), it can be verified that
\begin{align}
\partial S \bigcap \mathcal{C}(u,\theta(\epsilon))
=& \{x\in \partial S: u^Tx \ge  \|x\| \|u\|\delta(\epsilon) \} \label{eqn:partSdminuSep} \\
\subseteq & \{x\in \partial S: u^Tx \ge  \|u\|d_{\min}(u,S,\epsilon)\delta(\epsilon) \} \nonumber
\end{align}
where $d_{\min}(v,S,\epsilon)$ is defined as in (\ref{eqn:dminvS}). Observe that $\textrm{conv}(\{x\in \partial S: u^Tx \ge \|u\|d_{\min}(u,S,\epsilon)\delta(\epsilon) \}) = \{x\in S: u^Tx \ge \|u\|d_{\min}(u,S,\epsilon)\delta(\epsilon)\}$. This, together with (\ref{eqn:partSdminuSep}), implies that 
$
\sup_{0 \le \alpha \le 1} \{\alpha:  \alpha u \not\in  \textrm{conv}(\partial S \bigcap \mathcal{C}(u,\theta(\epsilon)))) \} 
\ge   \delta(\epsilon)d_{\min}(u,S,\epsilon)/\|u\|.
$
Finally, we arrive at
$
\gamma(u,S,\epsilon) \ge \delta(\epsilon)d_{\min}(u,S,\epsilon)/\|u\|,
$
which implies that $\gamma_{\min}(S,\epsilon) \ge \underline{\gamma}(S,\epsilon)$.
This completes the proof. $\Box$

\bibliographystyle{unsrt}
\bibliography{Reference}

\end{document}